\documentclass[amsmath,showpacs,twocolumn,aps,prl]{revtex4}
\usepackage{bm}
\usepackage{graphics}
\begin{document}
\title{Effect of electron-electron interactions  on the conductivity of clean graphene}
\author{E.G.~Mishchenko}
\affiliation{Department of Physics, University of Utah, Salt Lake
City, Utah 84112, USA}
\begin{abstract}

Minimal conductivity of a single undoped graphene layer  is known to
be of the order of the conductance quantum, independent of the
electron velocity. We show that this universality does not survive
electron-electron interaction which results in the non-trivial
frequency dependence. We begin with analyzing the perturbation
theory in the interaction parameter $g$ for the electron self-energy
and observe the failure of the random-phase approximation. The
optical conductivity is then derived from the quantum kinetic
equation and the exact result is obtained in the limit when $g \ll 1
\ll g |\ln\omega|$.

\end{abstract}


\maketitle

{\it \underline{Introduction.}} Recent experiments on  transport in
graphene layers \cite{Geim,Kim,Her} have validated extensive
theoretical efforts directed at  understanding of various properties
of two-dimensional Dirac fermions. A lot of these efforts are
devoted to the zero temperature dc conductivity which has a
universal value of the order of the conductance quantum
\cite{fradkin,lee}. Such a value is not unexpected from the
dimension analysis  since the intrinsic (undoped) graphene lacks a
characteristic momentum scale. Two different minimal conductivities
are conventionally defined. The dc limit of an ac conductivity in a
clean graphene ($\tau^{-1}=0$, $\omega \to 0$) was shown to be
$\sigma = e^2/4\hbar$ \cite{ludwig}. Another possible definition of
a strict dc limit of impure graphene ($\omega = 0$, $\tau^{-1} \to
0$) gives a different, but numerically close value $\bar{\sigma} =
2e^2/\pi^2\hbar$ \cite{ludwig}. Recent calculations have largely
confirmed \cite{gusynin,kats,jakub,cserti,nomura,falk,peres} the
results of Ref.~\onlinecite{ludwig}, while others obtained different
values \cite{ziegler}. Conductivity in a bilayer graphene has also
been a subject of close theoretical attention
\cite{falko,extortion,ando,katsnel}.

It is noteworthy that the minimal conductivity $\sigma$ is very much
analogous to the optical conductivity of a two-dimensional electron
system with spin-orbit-split band structure, where it is due to the
``chiral resonance'', and the corresponding value $e^2/16\hbar$
\cite{MCE,MH} is exactly $4$ times smaller than $\sigma$ (which is
the degree of spin-nodal degeneracy in graphene). This analogy is
due to similarity in the chiral properties of the eigenstates in the
two systems.

In the present paper we demonstrate that the notable universality of
the values of $\sigma$ and $\bar{\sigma}$ does not hold in  the
presence of electron-electron interactions, which result in a strong
frequency dependence of the conductivity. Here, we concentrate on
the optical conductivity $\sigma(\omega)$ in the strict
disorder-free graphene ($\tau^{-1} = 0$) and show that the optical
conductivity is actually {\it suppressed} by interactions in
comparison with its ``universal'' value.

Single intrinsic 2D graphene layer is described by the chiral
Hamiltonian,
\begin{equation}
\label{ham} H=v_0\sum_{i \bf p} ~ \hat c^{i\dagger}_{{\bf p}} \hat
{\bm \sigma}\cdot {\bf p} ~\hat c^{i}_{\bf p}+\frac{1}{2}\sum_{ij
\bf pkq} \hat c^{i\dagger}_{ {\bf p-q}}\hat c^{j\dagger}_{{\bf k+q}}
V_{\bf q} \hat c^j_{\bf k} \hat c^i_{{\bf p}},
\end{equation}
where ``hats'' denote operators in pseudo-spin space ($\hat{\bm
\sigma}$ represents the usual set of Pauli matrices), the sum over
Latin indices is taken over two nodal points and two (true) spin
directions. The interaction potential is $V_{\bf q}=2\pi e^2/\kappa
|{\bf q}|$, with $\kappa$ being the dielectric constant of a
substrate, $v_0$ is the ``bare'' electron velocity. Hereinafter we
denote, $\sum_{\bf p} \equiv \int {d^2p}/{(2\pi)^2}$, and set
$\hbar=1$ throughout the text, except in the final result
(\ref{final_answer}).

 Let us begin with the perturbation theory, in powers of the
dimensionless interaction constant $g=e^2/\kappa v_0$, for the
electron self-energy (at $T=0$). We find that the random phase
 approximation (RPA) {\it fails} for the system described by the Hamiltonian (\ref{ham}),
 as the non-RPA contributions
 are generally not small.

 {\it \underline{Perturbation theory.}} The first and second order interaction corrections to the electron self-energy
 are shown in
 Fig.~1. The solid line corresponds to the electron
 Green's function
 \begin{equation}
 \label{subband}
 \hat G_{\epsilon {\bf p}}=\frac{1}{\epsilon-v_0p \hat\sigma_{\bf p} +i\eta \text{sgn} {\epsilon}}=
 \frac{1}{2} \sum_{\beta=\pm 1} \frac{1+\beta \hat \sigma_{\bf
 p}}{\epsilon-\beta(
 v_0p-i\eta)},
 \end{equation}
 where $\hat\sigma_{\bf p}=   \hat{\bm \sigma}\cdot {\bf n}$ is the
 projection of the pseudo-spin operator onto the direction of the
 electron momentum ${\bf n}={\bf p}/p$.
The first-order contribution (Fig.~1a) is independent of the energy
variable,
       \begin{equation}
\hat \Sigma^a_{\bf p}=i\sum_{\bf p'}\int \frac{d\epsilon}{2\pi}
V_{\bf p-p'} \hat G_{\epsilon {\bf p'}}  =\frac{1}{2}~\hat{
\sigma}_{\bf p} \sum_{ \bf p'} V_{\bf p-p'} ({\bf n}\cdot{\bf n'}).
\end{equation}
The integral here diverges logarithmically at $p'\gg p$. This
divergence is cut-off at the inverse lattice spacing, ${\cal K}$,
leading to the logarithmic correction,
\begin{equation}
\label{1a} \hat \Sigma^a_{\bf p}=p~\hat{ \sigma}_{\bf
p}\frac{e^2}{4\kappa}\ln({\cal K}/p).
\end{equation}
Such logarithmic renormalization of the electron velocity was first
discussed in Ref.~\onlinecite{Gonzales} within the scope of the
renormalization group approach.

\begin{figure}[h]
\resizebox{.28\textwidth}{!}{\includegraphics{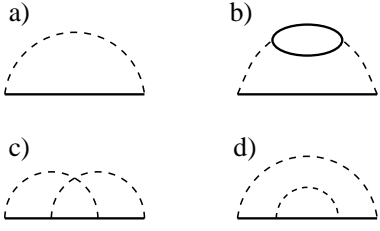}}
\caption{First (a) and second (b,c,d) order corrections to the
electron self-energy. Solid line represents the electron Green's
function (\ref{subband}), dashed line stands for the bare
interaction potential $V_{\bf q}$. Closed loop implies summation
over nodal and spin degeneracy.}
\end{figure}
The second order correction consists of three contributions,
Fig.~1b, c, and d, of which the last one is identically zero. This
fact can be understood upon noticing that the diagram 1d is
responsible for the renormalization of the Fermi level \cite{Mahan}
and must be absent in clean graphene due to its electron-hole
symmetry. The other two graphs give non-zero contributions. The
RPA-type diagram Fig.~1b gives
       \begin{equation}
       \label{sigmab}
\hat \Sigma^b_{\epsilon \bf p}=4i\sum_{\bf q}\int
\frac{d\omega}{2\pi} V^2_{\bf q} \hat G_{\epsilon-\omega, {\bf p-q}}
\Pi(\omega,{q}),
\end{equation}
where the coefficient $4$ comes from the summation over the nodal
points and (real) spin states in the internal loop which represents
the polarization operator
 \begin{equation}
\Pi(\omega,{q})=-i\text{Tr}\sum_{\bf p'}\int \frac{d\epsilon'}{2\pi}
\hat G_{\epsilon'+\omega,{\bf p'+q}} \hat G_{\epsilon' {\bf p'}},
\end{equation}
here trace is taken in the pseudo-spin space. To evaluate the
integrals it is convenient to use the following integral
representation for the polarization operator,
 \begin{equation}
 \label{pol}
\Pi(\omega,q)=\frac{ q^3v_0}{8\pi}\int_0^\infty
\frac{dx}{\omega^2-q^2v_0^2(1+x^2)+i\eta}.
\end{equation}
The integral over frequency in Eq.~(\ref{sigmab}) can now be easily
calculated. Restricting for simplicity to the $\epsilon=0$ case here
\cite{OK}, we obtain,
\begin{equation}
\hat \Sigma^b_{\bf p}=- \frac{\pi e^4}{\kappa v_0}\sum_{\bf q}
\int_0^\infty\frac{dx}{\sqrt{1+x^2}} \frac{\hat \sigma_{\bf
 p-q}}{q\sqrt{1+x^2} +|{\bf p-q}|}.
\end{equation}
The main (logarithmic) contribution to this integral comes from
large values of transferred momentum $q$. Expanding to the linear
order in $p/q$, and evaluating first the integral over the angle
between ${\bf p}$ and ${\bf q}$, then the integral over $dx$ and
finally over the absolute value of $q$, we arrive at
\begin{equation}
\label{1b} \hat \Sigma^b_{\bf p}=-p~\hat \sigma_{\bf
 p} \frac{ e^4}{6\kappa v_0}\ln({\cal K}/p).
\end{equation}
The remaining contribution Fig.~1c is given by (again,
$\epsilon=0$),
\begin{eqnarray}
 \hat \Sigma^c_{\bf p}&=&\frac{1}{4v_0} \sum_{\bf p'
p''}V_{\bf p-p'} V_{\bf p-p''}\nonumber\\ &\times &
\frac{\hat\sigma_{\bf p'}+\hat\sigma_{\bf p''}-\hat\sigma_{\bf
p'+p''-p}-\hat\sigma_{\bf p'}\hat\sigma_{\bf
p'+p''-p}\hat\sigma_{\bf p''}}{p'+p''+|{\bf p'+p''-p}|}.~~~
\end{eqnarray}
The leading logarithmic contribution into this expression comes from
large $p',p'' \gg p$. To this leading order, it is sufficient  to
expand the integrand to the linear power in ${\bf p}$. As the
relevant terms appear from the interaction potentials as well as
from the expansion of  both the numerator and denominator, the
calculations are too cumbersome to be presented here. Finally,
\begin{equation}
\label{1c} \hat \Sigma^c_{\bf p}=p~\hat \sigma_{\bf
 p} \frac{ e^4}{\kappa v_0}\left(-\frac{2}{3}+\ln 2 \right)\ln({\cal K}/p).
\end{equation}
The expressions (\ref{1a},\ref{1b},\ref{1c}) determine the
remormalization of the electron velocity up to the second order in
the electron-electron interaction,
\begin{equation}
\label{velocity} \frac{v_p}{v_0}=1+\left[ \frac{g}{4}
-g^2\left(\frac{5}{6}-\ln 2 \right)+O(g^3)\right]\ln({\cal K}/p).
\end{equation}
Two important conclusions can be drawn from this result.

 (i) The contribution (\ref{1c}) from the non-RPA diagram
with intersecting interaction lines {\it is not parametrically
small} compared with the RPA term (\ref{1b}). Resulting in an
overestimation of about $\sim 20\%$ in the second order, the neglect
of non-RPA corrections to the electron self-energy becomes an
uncontrollable approximation in the higher orders in $g$ for an
undoped graphene. Our findings, thus, do not support the conjecture
of Refs.~\onlinecite{Sarma} (where it was used for the calculation
of the electron lifetime) that RPA is exact approximation in the
limit $g \ll 1$. Experimentally, the value of the velocity is
$v_0=1.1\times 10^6$ m/s \cite{Kim}. For a typical dielectric
substrate $\kappa \approx 6$, which yields $g \approx 0.3$,
indicating that $g \ll 1$ is a reasonable approximation.

 (ii) It is essential that the higher orders {\it do not produce}
higher powers of $\ln({\cal K}/p)$. This is most simply verified for
the RPA-terms \cite{Gonzales}. On the other hand, the diagrams with
intersecting interaction lines are not more singular than the RPA
ones as is clear from the power counting (but also not
parametrically small than RPA terms). As the consequence, in the
limit $g\ll 1$ it is sufficient to restrict to the lowest order of
the perturbation theory. Still, the product $g\ln({\cal K}/p)$ {\it
can be arbitrarily large} without violating the perturbation
expansion.

{\it \underline{Kinetic equation}}. We now apply the obtained
understanding to the analysis of the ac conductivity
$\sigma(\omega)$. Free-electron conductivity follows directly from
the polarization operator (\ref{pol}) and the particle conservation
condition. Taking into account nodal and spin degeneracy,
\begin{equation}
\sigma(\omega) = 4\lim_{q \to 0} \frac{ie^2\omega}{q^2}
\Pi(\omega,q) = \frac{e^2}{4}.
\end{equation}
In order to calculate the homogeneous ac conductivity in the
presence of electron-electron interactions we apply a quantum
kinetic equation. This will allow us to take into account the
leading terms $\sim g\ln({\cal K}/p)$ in the exact  way (but still
assuming $g \ll 1$). The density matrix $\hat f_{\bf
 p}$ should be defined as a matrix in a pseudo-spin space,
$ f^{\alpha \beta}_{\bf p} = \langle\langle c^{\beta \dagger}_{\bf
p} c^\alpha_{\bf p}\rangle \rangle,$  where $\alpha$ and $\beta$ are
the pseudo-spin indices. Using the equations of motion for the
operators $\hat c_{\bf p}$ determined by the Hamiltonian
(\ref{ham}), and assuming a homogeneous in space (but
time-dependent) external electric field ${\bf E}$, it is
straightforward to write a closed equation for the density matrix
$\hat f_{\bf p}$ to the linear order in the electron-electron
interactions,
\begin{equation}
\label{kinetic} \frac{\partial \hat f_{\bf p}}{\partial t}+ivp
[\hat{ \sigma}_{\bf p},\hat f_{\bf p}] +e{\bf E}\cdot\frac{\partial
\hat f_{\bf p}}{\partial \bf p}=i \sum_{\bf p'} V_{\bf p-p'} [\hat
f_{\bf p'},\hat f_{\bf p}].
\end{equation}

We emphasize that kinetic equation for graphene (\ref{kinetic}) does
contain the terms linear in the interaction. This is in contrast to
the conventional system with parabolic dispersion for which
non-trivial contributions (collision integral) appear in the second
order in $V_{\bf q}$ \cite{RS}. This difference is due to the chiral
structure of the electron eigenstates in graphene.
 Kinetic equation needs to be solved to the linear order in the
 applied
electric field, $\hat f_{\bf p}=\hat f^{(0)}_{\bf p}+\hat f_{\bf
p}^{(1)}$, around the equilibrium distribution  $\hat f^{(0)}_{\bf
p}=1-\hat{ \sigma}_{\bf p}/2$ (at zero temperature). The
contribution resulting from the term $[\hat f^{(0)}_{\bf p'},\hat
f^{(1)}_{\bf p}]$ represents simply the electron self-energy
$\Sigma_{\bf p}^a$ to the first order in interaction. The remaining
term $[\hat f^{(1)}_{\bf p'},\hat f^{(0)}_{\bf p}]$ describes vertex
corrections. For the periodic electric field the linearized kinetic
equation takes the form
\begin{eqnarray}
\label{kinur_homogen} -i\omega \hat  f_{\bf p}^{(1)}+i v_p p
~[\hat{\sigma}_{\bf p},\hat  f_{\bf p}^{(1)}] = \frac{e}{2p} ({\bf
E}\times {\bf n})\cdot(\hat{\bm \sigma}\times {\bf
n})\nonumber\\-\frac{i}{2}\sum_{\bf p'}V_{\bf p-p'} [\hat
f^{(1)}_{\bf p'},\hat \sigma_{\bf p}],
\end{eqnarray}
where the renormalized velocity $v_p$ is given by the expression
(\ref{velocity}) taken to the first order in $g$. With the following
substitution,
\begin{equation}
\hat f_{\bf p}^{(1)}= i ({\bf E}\times {\bf n})\cdot (\hat {\bm
\sigma}\times {\bf n})A(p)+{\bf E}\cdot (\hat {\bm \sigma}\times
{\bf n}) B(p),
\end{equation}
the matrix integral equation (\ref{kinur_homogen}) reduces to a pair
of coupled scalar integral equations,
\begin{eqnarray}
\label{system}
 \omega A(p) +2 v_p p B(p)&=&\frac{e}{2p}+\sum_{\bf
p'}
V_{\bf p-p'}B(p') \cos\theta_{pp'} \nonumber\\
2 v_p p A(p) +\omega B(p) &=&\sum_{\bf p'}V_{\bf p-p'}A(p')
\cos^2\theta_{pp'},
\end{eqnarray}
where $\theta_{pp'}$ is the angle between the vectors ${\bf p}$ and
${\bf p'}$. The system of equations (\ref{system}) can be solved by
means of the consecutive approximations in $V_{\bf p-p'}$. Fig.~2
illustrates  diagrammatically the meaning of these consecutive
orders. In the limit $g\ll 1$ it is sufficient to stop at the first
order term (see the discussion at the end of this section),
\begin{figure}[h]
\resizebox{.47\textwidth}{!}{\includegraphics{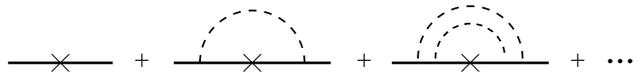}}
\caption{Graphic representation of the density matrix given by the
integral equation (\ref{kinur_homogen}). The dashed line represents
the interaction potential $V_{\bf q}$, the solid line stands for the
electron propagator corrected by the self-energy $\Sigma_{\bf p}^a$.
The cross denotes the applied external field. The first order term
brings a logarithmic contribution into the optical conductivity,
$\propto \int dp/v_pp$. The higher-order terms contain higher power
of the small coupling constant $g$ but do not result in higher
powers of logarithms.}
\end{figure}
\begin{eqnarray}
\label{solution_a} A(p)&=&\frac{e}{2p}\frac{\omega}{\omega^2-4 v_p^2
p^2}\nonumber\\ &-& e\omega \sum_{\bf p'}V_{\bf p-p'} \frac{  v_p p
\cos^2\theta_{pp'} +  v_{p'} p'\cos\theta_{ pp'} }{p'(\omega^2-4
v^2_pp^2)(\omega^2-4  v^2_{p'}p'^2)}.~~~~~~
\end{eqnarray}
Here, as usual, the frequency should be understood as posessing the
infinitesimally small positive imaginary part, $\omega \to
\omega+i\eta$. Knowledge of the function $A(p)$ allows to determine
the electric current, ${\bf j} = 4e \text{Tr} \sum_{\bf p} v_0 \hat
f_{\bf p} \hat {\bm \sigma} = \sigma(\omega) {\bf E}$, which yields
the conductivity
\begin{equation}
\sigma(\omega) =\frac{2ei}{\pi}~ \int_0^\infty d{p}~ v_0 p A(p).
\end{equation}
The real part of the conductivity is due to the zeros of the
denominators of Eq.~(\ref{solution_a}). In the second term in
Eq.~(\ref{solution_a}) the leading logarithmic contribution comes
from the imaginary part that arises due to the second singularity
($|\omega|=2v_{p'}p'$). It is important that the second order term
in the expansion Eq.~(\ref{solution_a}), which is not explicitly
written, {\it does not} yield any $g^2\ln^2p$ contribution, being
only $\sim g^2 \ln{p}$, and, thus, by virtue of $g \ll 1$ resulting
only in a small correction to the first term, similar to the
situation already encountered in the self-energy, cf.\
Eq.~(\ref{velocity}). Upon extracting this imaginary part, one
should first evaluate the integral over the angle $\theta_{pp'}$ and
then over the momentum $p$. Straightforward calculation leads to the
low-frequency optical conductivity
\begin{equation}
\label{final_answer} \sigma'(\omega) = \frac{e^2}{4\hbar}
\frac{v_0}{v_\omega} \left[1+ \ln
\left(\frac{v_\omega}{v_0}\right)\right],
\end{equation}
where $v_\omega =v_0[1+{g}\ln \left(\frac{v_0 \cal
K}{|\omega|}\right)/4]$ is the renormalized electron velocity,
corresponding to frequency $\omega$. In deriving
Eq.~(\ref{final_answer}) it is utilized that $
\int_{\omega/v_0}^{\cal K}{dp}/{v_p p}=4\ln
\left(\frac{v_\omega}{v_0}\right)/{gv_0}$.

Eq.~(\ref{final_answer}) is the main result of this paper. The
conductivity depends {\it only} on the combination of the frequency
and the coupling constant, ${g}\ln \left(\frac{v_0 \cal
K}{|\omega|}\right)$. We illustrate the dependence of the optical
conductivity on the frequency for different values of the coupling
constant $g$ in Fig.~3. When frequency increases the optical
conductivity approaches its value $e^2/4\hbar$, but at lower
frequencies its magnitude is significantly reduced.
\begin{figure}[h]
\resizebox{.37\textwidth}{!}{\includegraphics{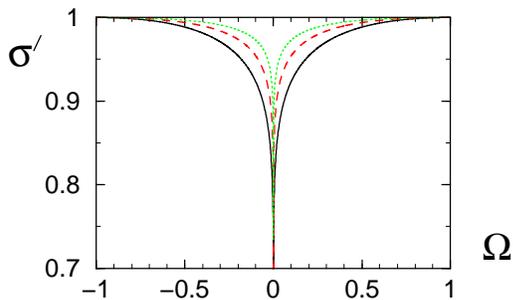}}
\caption{(Color online) The dependence (\ref{final_answer}) of the
real part of the  optical conductivity (measured in units of the
non-interacting conductivity $e^2/4\hbar$ on the dimensionless
frequency $\Omega=\omega/v_0 {\cal K}$ for different values of the
coupling constant: $g=0.5$ (green dotted line), $g=0.7$ (red dashed
line), $g=1.0$ (black solid line). The suppression of the
conductivity is pronounced in the low-frequency domain.}
\end{figure}

Restricting the perturbation expansion (\ref{solution_a}) to the
lowest order vertex correction requires some justification. In
particular, one has to make sure that this expansion does not lead
to higher-order logarithms in the discarded terms, i.e. that no
x-ray (Mahan) singularity \cite{Mahan} takes place. Indeed, the
second order contribution into $A(p)$ can be easily derived and
analyzed. The corresponding expression is cumbersome  but the
conclusion is straightforward. The higher order terms still feature
single-logarithmic divergencies but arrive with higher powers of
$g$. The situation here is similar to the one already encountered
above in the calculation of the electron self-energy, see
Eq.~(\ref{velocity}).

{\it \underline{Summary and conclusions}}. Optical conductivity in
undoped graphene is the result of the ``chiral'' resonance, i.e.
resonant creation of an electron-hole pair. Chiral structure of the
Hamiltonian leads to non-zero matrix elements of the velocity
operator for the inter-subband transitions, resulting in a finite
conductivity in a homogeneous electric field \cite{Drude}.

Effects of electron-electron interaction are two-fold. First, the
logarithmic renormalization of the electron velocity at low energies
leads to the {\it decrease} in the density of states, which {\it
suppresses} the probability of inter-subband transitions. This
effect is accounted for by the electron self-energy and reveals in
the appearance of the factor $v_0/v_\omega$ in the expression
(\ref{final_answer}). Second, the excited electron and hole interact
in the final state. This interaction is attractive and results in
the relative  {\it enhancement} of the optical conductivity. Such
final-state interactions are analogous to the excitonic effect in
conventional semiconductors and are accounted for by the vertex
corrections, Fig.~2, yielding the factor $1+ \ln
\left(\frac{v_\omega}{v_0}\right)$ into Eq.~(\ref{final_answer}).
Utilization of the quantum kinetic equations allows to consider both
these effects on equal footing and account for the leading
logarithmic terms in the limit when $g\ll 1 \ll g\ln \left(\frac{v_0
\cal K}{|\omega|}\right)$. Note that in order to be able to study
homogeneous conductivity one should be able to neglect charge
accumulation at the boundaries and associated with it electric
field. It is easy to estimate that the system size should be larger
enough, $L \gg gv_0/\omega$.

Useful discussions with M. Raikh and O. Starykh are gratefully
acknowledged. The author is also thankful to V. Zyuzin for his help
in numerical verification of some integral identities related to the
calculations of the electron self-energy.
 This work was supported by DOE, Office of Basic Energy
Sciences, Award No.~DE-FG02-06ER46313.


\begin{thebibliography}{50}
\bibitem{Geim} K. S. Novoselov, A. K. Geim, S. V. Morozov, D. Jiang, Y.
Zhang, S. V. Dubonos, I. V. Grigorieva, and A. A. Firsov, Science
{\bf 306}, 666 (2004); K. S. Novoselov, A. K. Geim, S. V. Morozov,
D. Jiang, M. I. Katsnelson, I. V. Grigorieva, S. V. Dubonos, and A.
A. Firsov, Nature {\bf 438}, 197 (2005).

\bibitem{Kim} Y. Zhang, J. P. Small,
M. E. S. Amori, and P. Kim, Phys. Rev. Lett. {\bf 94}, 176803
(2005); Y. Zhang, Y.-W. Tan, H. L. Stormer, and P. Kim, Nature {\bf
438}, 201 (2005).

\bibitem{Her} X. Wu, X. Li, Zh. Song, C. Berger, W.A. de Heer,
Phys. Rev. Lett. {\bf 98}, 136801 (2007).

\bibitem{fradkin} E. Fradkin, Phys. Rev. B {\bf 33}, 3263 (1986).

\bibitem{lee}
P. A. Lee, Phys. Rev. Lett. {\bf 71}, 1887 (1993).
\bibitem{ludwig}
A. W. W. Ludwig, M. P. A. Fisher, R. Shankar, and G. Grinstein,
Phys. Rev. B {\bf 50}, 7526 (1994).





\bibitem{gusynin} V. P. Gusynin and S. G. Sharapov, Phys. Rev. Lett. {\bf 95}, 146801
(2005).


\bibitem{kats} M. I. Katsnelson, Eur. J. Phys. B {\bf 51}, 157 (2006).

\bibitem{jakub} J. Tworzydlo, B. Trauzettel, M. Titov, A. Rycerz, C.W.J.
Beenakker, Phys. Rev. Lett. {\bf 96}, 246802 (2006).

\bibitem{cserti} J.~Cserti, Phys. Rev. B {\bf 75}, 033405 (2007).




\bibitem{nomura} K. Nomura and A. H. MacDonald, cond-mat/0606589.

\bibitem{falk} L. A. Falkovsky and A. A. Varlamov, cond-mat/0606800.

\bibitem{peres} N. M. R. Peres, F. Guinea, and A. H. Castro Neto, Phys. Rev. B
{\bf 73}, 125411 (2006).

\bibitem{ziegler}
K. Ziegler, Phys. Rev. Lett. {\bf 97}, 266802 (2006).

\bibitem{MCE} L.I. Magarill, A.V. Chaplik and M.V. \'Entin, JETP
{\bf 92}, 153 (2001).
\bibitem{MH} E.G.~Mishchenko and B.I.~Halperin, Phys.\ Rev.\ B {\bf
68}, 045317 (2003).

\bibitem{falko}
E. McCann and V. I. Fal'ko, Phys. Rev. Lett. 96, 086805 (2006).

\bibitem{extortion}  J. Nilsson, A. H. Castro Neto, F. Guinea, and N. M. R.
Peres, Phys. Rev. Lett. {\bf 97}, 266801 (2006).


\bibitem{ando} M. Koshino and T. Ando, Phys. Rev. B {\bf 73}, 245403 (2006).

\bibitem{katsnel} M. I. Katsnelson, Eur. Phys. J. B {\bf 52}, 151-153 (2006).

\bibitem{Gonzales} J. Gonz\'alez, F. Guinea, and M.A.H. Vozmediano,
Phys. Rev. B {\bf 59}, R2474 (1999).

\bibitem{Mahan} G. Mahan, {\it Many-particle physics} (Kluwer, New
York, 1981).

\bibitem{OK} Conclusions similar to ones obtained in this paper
can be reached for the renormalization of the energy term in the
denominator of the Green's function (\ref{subband}). Again,
single-logarithmic terms are encountered in higher orders of the
perturbation theory (see also Ref.~\onlinecite{Gonzales} where RPA
terms were considered). Such terms will not be important for the
calculation of the electric conductivity, which requires the
knowledge of kinetic equation (\ref{kinetic}) for the density matrix
at coinciding times only.

\bibitem{Sarma} S. Das Sarma, E.H. Hwang, and W-K. Tse,
cond-mat/0610581; E.H. Hwang, B.Yu-K. Hu, and S. Das Sarma,
cond-mat/0612345.

\bibitem{RS} J.~Rammer and H.~Smith, Rev.\ Mod.\ Phys. {\bf 58},
323 (1986).

\bibitem{Drude} At finite temperatures intra-subband transitions lead
to the temperature-dependent Drude peak $\sigma' \propto T\delta
(\omega)$. We do not discuss $T\ne 0$ case in this paper.

\end{thebibliography}
\end{document}